\documentclass{article}

\usepackage{arxiv}

\usepackage[utf8]{inputenc} % allow utf-8 input
\usepackage[T1]{fontenc}    % use 8-bit T1 fonts
\usepackage{hyperref}       % hyperlinks
\usepackage{url}            % simple URL typesetting
\usepackage{booktabs}       % professional-quality tables
\usepackage{amsfonts}       % blackboard math symbols
\usepackage{nicefrac}       % compact symbols for 1/2, etc.
\usepackage{microtype}      % microtypography
\usepackage{lipsum}
\usepackage{mparhack}
\usepackage{xcolor}
\usepackage{multirow}
\usepackage{nicefrac}
\usepackage{subcaption} 
\usepackage{array}
\usepackage{graphicx}
\newcolumntype{P}[1]{>{\centering\arraybackslash}p{#1}}

\begin{document}

\title{E-Gotsky: Sequencing Content using the Zone of Proximal Development}
\author{
        Oded Vainas\thanks{This work was done while the author was with Microsoft}\\
        SimilarWeb\\
        Israel\\
    \And
        Ori Bar-Ilan\footnotemark[1]\\
        Zeitgold\\
        Israel\\
    \And
        Yossi Ben-David\\
        Microsoft\\
        Israel\\
    \And
        Ran Gilad-Bachrach\\
        Microsoft\\
        Israel\\
    \And
        Galit Lukin\\
        MIT\\
        MA, USA
    \AND
        Meitar Ronen\\
        Microsoft\\
        Israel\\
     \And
        Roi Shillo\\
        Ben-Gurion University\\
        Israel\\
    \And
        Daniel Sitton\\
        Microsoft\\
        Israel\\
}

\maketitle
\begin{abstract}
Vygotsky's notions of Zone of Proximal Development and Dynamic Assessment emphasize the importance of personalized learning that adapts to the needs and abilities of the learners and enables more efficient learning. In this work we introduce a novel adaptive learning engine called E-gostky that builds on these concepts to personalize the learning path within an e-learning system. E-gostky uses machine learning techniques to select the next content item that will challenge the student but will not be overwhelming, keeping students in their Zone of Proximal Development.

To evaluate the system, we conducted an experiment where hundreds of students from several different elementary schools used our engine to learn fractions for five months. Our results show that using E-gostky can significantly reduce the time required to reach similar mastery. Specifically, in our experiment, it took students who were using the adaptive learning engine $17\%$ less time to reach a similar level of mastery as of those who didn't. Moreover, students made greater efforts to find the correct answer rather than guessing and class teachers reported that even students with learning disabilities showed higher engagement.

\end{abstract}

\keywords{Personalized learning path, e-learning, Adaptive learning, Zone of Proximal Development} 

\section{Introduction}
Today, ``Teaching for the average student" is still the prevailing strategy, a strategy which keeps struggling students behind and is not utilizing the time effectively for other students which are left disengaged~\cite{darling2006no}. Creating a personalized learning environment is challenging in traditional classrooms, where the teacher is outnumbered by the students. However, modern technology allows supporting the teacher in ways that can allow effective personalized learning~\cite{darling2006no}.

The notion of personalized learning is far from new. Lev Vygotsky conceptualized the idea of \textit{Zone of Proximal Development} (ZPD) in the early 1930s, a well-known and vastly researched concept of educational psychology, and laid the foundations for the concept of personalized learning~\cite{vygotsky1980mind, chaiklin2003zone}. In this work we present a quantitative interpretation of the ZPD concept, translating it to a data driven algorithm that supports the customization of a given education program and enables the delivery of personalized learning content. 

ZPD is defined as ``the distance between the actual developmental level as determined by independent problem solving and the level of potential development as determined through problem-solving under adult guidance, or in collaboration with more capable peers" \cite[pg. 86]{vygotsky1980mind}. In other words, it refers to what a learner can do with assistance but cannot do independently. According to Vygotsky, concrete growth can only occur in the ZPD, and the learning is most effective when the support is matched to the needs of the learner. 

Many e-learning systems contain a large set of exercises\footnote{The terms ``exercise" includes also other activities such as watching instructional videos.} such that the students moves in a linear order between exercises. Providers of such content face a dilemma: on one hand, the exercises should progress slowly to allow struggling students to master topics before new concepts are introduced. However, such slow progress does not serve well the needs of other students who may find themselves bored and disengage. Therefore, content served in a non-adaptive manner will under-challenge some students, while over-challenging others. In view of Vygotsky's theory, it can be said that the problem stems from the fact that each student has a different ZPD at a given point in time.

The method proposed here tries to mitigate this problem. Whenever a student solves an exercise, an adaptive learning engine, named E-gostky, can move forward or backward in the content to find the next exercise that is not too easy and not too hard. For example, it can skip over some exercises for a student that already demonstrated mastery of the topic while skipping over harder than normal exercises (``bonus exercises") for students that are still struggling. Therefore, E-gostky tries to make sure that the next exercise served will keep students in their ZPD.

Identifying whether a task will keep a student within her ZPD is not an easy task and it has been researched extensively~\cite{lloyd1999lev, chaiklin2003zone}. To this extent, we use another core concept of Vygotsky's theory: Dynamic Assessment (DA). The goal of DA is to assess the potential for learning, rather than a static level of achievement, by prompting students to use their minds and assistance in problem solving. By dynamically evaluating students' progress, we strive to adjust the content to their learning potential at any given moment. In other words, students are not categorized into groups. Instead, adaptation is done based on the learning potential of the student at the current time and on the current topic.

To test our method, we used content that was developed by pedagogical experts to serve elementary school students while learning fractions. We conducted a randomized controlled trial where this content, in its original form was used as a baseline method and compared it to the same content when it is augmented by the E-gostky adaptive learning engine. The two approaches were compared 
on different sections of the learning program. We found that students in both groups achieved similar mastery levels but the students on the adaptive track required significantly less time to reach this level: on average, students spent $~17\%$ less time interacting with the delivered content - a time that can be used by the teacher for other educational activities.
 
 The qualitative feedback provided by teachers suggests that students using E-gostky were more engaged and eager to learn. One of the teachers reported that ``using the skipping engine (E-gostky) was interpreted as positive feedback that the student wanted to receive more of, so it motivated them to learn, think longer before answering, and use the aids to answer correctly".
 % I believe this is true for the e-learning platform in general, and not specifically to our approach - Meitar
 %Other teachers reported on the benefits the approach provided to struggling students who were more engaged and got better support.

\section{Background and Related Work}

The Zone of Proximal Development (ZPD) plays a key role in the method we propose. The ZPD was introduced by Vygotsky as a part of a general analysis of the development of children. Vygotsky proposed a model of child development where each period is characterized by a set of relations among psychological functions (such as perception, speech, thinking), and the transition from one age period to another depends on a qualitative change in that set of relations \cite{chaiklin2003zone}. In this model, learning is a function not only of the child's qualities but also of her relationships with the environment.

The ZPD in Vygotsky's theory is used to identify which psychological functions (and related social interactions) are needed for transitioning to the next age period, and to assess the current status of the child's maturing functions~\cite{chaiklin2003zone}. Since performance in these psychological functions depends on social interactions, the assessment procedures should have dynamic, interactive nature. Hence, the ZPD provides a framework to evaluate learner's abilities, which is known as Dynamic Assessment (DA)~\cite{chaiklin2003zone, lantolf2011dynamic}. While in traditional approaches to assessment the main concern is current existing skills, DA is focused on measuring the learning potential for future development~\cite{lantolf2011dynamic}. A dynamic test is based on teacher assistance, where guidance, feedback and adaptive delivery of assistance are embedded in the evaluation procedure itself~\cite{chaiklin2003zone}. Rather than information about past functioning and existing skills, dynamic approaches tend to be equally interested in estimating the learner's cognitive and meta-cognitive strategies, their responsiveness to assistance, and their ability to transfer skills that were learned with assistance to subsequent unassisted situations~\cite{lidz2006use}.

While originating from a developmental theory, the ZPD and DA are widely adopted in educational for various areas such as reading, writing, mathematics, second-language learning and musical teaching, with diverse groups of students ranging from preschoolers to adults~\cite{chaiklin2003zone, lantolf2011dynamic, lidz2006use, bosma2010teacher}. The benefits of implementing these concepts in educational applications have been demonstrated by many scholars. It has been showed that teaching in the ZPD increases motivation and results in higher performance~\cite{ tieso2005effects, topping1996effectiveness, walpole2007differentiated, dixon1995partner}.

Our work also relates to past research on using historical data to sequence content for students, and the work on ZPD for online adaptation of educational content. There are two basic types of adaptive behavior in educational systems, described by different authors and mostly called inner loop and outer loop~\cite{pelanek2017bayesian}. In the inner loop the focus is on a single multi-step problem while the outer loop focuses on a sequence of independent items. In this paper we are focusing on an approach for the outer loop as suggested by previous studies that argued that adaptive systems should focus on macro-adaptivity (outer loop) and that the micro-adaptivity should be the focus of the instructor~\cite{essa2016possible}.

The problem of sequencing educational content attracted many researchers, especially in the educational data mining community \cite{pelanek2017bayesian}.
For example, Pardon and Heffernan~\cite{pardos2009determining} inferred order over exercises presented to students by predicting their skill levels using Bayesian Knowledge Tracing (BKT)~\cite{corbett1994knowledge}. They showed the efficacy of their approach on simulated data as well as on a test set comprising random sequences of three exercises. Ben-David et al.~\cite{david2016sequencing} proposed a BKT based sequencing algorithm that uses knowledge tracing to model students' skill acquisition over time and sequences exercises based on their mastery level and predicted performance. The setting of this work is similar to the setting we study here, however, they use different method for sequencing which does not take the ZPD into account. Segal et al.~\cite{shani2014edurank, segal2018combining} introduced several sequencing algorithms. The first, EduRank, an algorithm that combines collaborative filtering with social choice theory to produce personalized learning sequences for students. The latter, MAPLE, an algorithm that combines difficulty ranking with Multi Armed Bandits. Clement et al.~\cite{clement2013multi} used experts' knowledge to bootstrap a Multi-Armed Bandit approach with models that rely on empirical estimation of the learning progress.

Several attempts have been made to utilize ZPD in an Intelligent Tutoring System and e-learning systems, for example to recommend on social activities~\cite{de2013strategy}. Salomon et al.~\cite{salomon1989computer} used the ZPD framework in a computerized Reader Partner software, and showed higher improvements in reading comprehension and in the quality of written essays for those who used the ZPD-based learning in comparison to a control group. Another way to formalize the ZPD was demonstrated in the AnimalWatch system, which teaches fractions in elementary schools~\cite{arroyo2000animalwatch, beal2010evaluation}. This approach presented content and hints according to the student model suggested by Beck et al.~\cite{beck1997using}. These approaches assume that the content can be dynamically generated in the desired difficulty level while the system we present in the following section evaluates student's performance and exercises simultaneously.

\section{Methods}\label{sec:methods}
The adaptive learning method we propose consists of two components: (i) \emph{Dynamic Assessment} - this component predicts student's performance in proposed exercises (ii) \emph{Sequencing Policy} - this component is a policy that given the predictions made using the Dynamic Assessment decides which exercises are in the current ZPD of the student and assigns the next exercise.

The Dynamic Assessment component predicts, for a student-exercise pair, two factors: (1) \emph{Time To Success (TTS)} - how much time will the student spend until she finds the correct solution to the exercise, and (2) \emph{Correct at First Attempt (CFA)} - the probability that the student will solve the exercise correctly in her first trial. These measures have been used in various studies as an estimate of student knowledge or performance~\cite{usingCFA, toscher2010collaborative, brinton2016mining, reckase2009multidimensional, pelanek2018exploring}. The Dynamic Assessment component uses machine learning to train two models, one for TTS and one for CFA using features such as:
\begin{enumerate}
    \item \textbf{Content Features:} features that are exercise-specific and derived from the performance of students that have attempted to solve this exercise. These features include the mean and standard deviation (STD) of CFA and TTS. These features are updated periodically as more students engage with the system. 
    \item \textbf{Student-Content Features:} for each student, the system keeps track of the mean and STD of the CFA and TTS of the student per each exercise type. That is, the rate in which the student answers correctly in the first attempt a multiple-choice exercise, the time to solve a bonus exercise and so on.
    \item \textbf{Student Online Features:} the mean and STD of the CFA and TTS in the latest exercises attempted by the student. For example, the CFA and TTS during the last 1, 3, 5, and 20 exercises. These features capture the current state of the student, both in terms of understanding the topic being taught and her readiness to learn.
\end{enumerate}
To reduce the complexity of the system, feature selection is used to select a subset of the above features to be used at run time.
The set of student online features represents the incorporation of Dynamic Assessment (DA) to enhance learning. We collect these measures repeatedly to estimate the status of the learner and her learning potential. Since the content is intertwined with interactive instruction, aids and feedback (further described in~\ref{sec:content}) we concur with Lidz and Elliot's summary of DA: ``The essential characteristics of DA are that they are interactive, open ended, and generate information about the responsiveness of the learner to intervention"~\cite{lidz2000application}.

\begin{figure}
\begin{center}
  \begin{subfigure}[b]{0.23\textwidth}
    \includegraphics[width=\textwidth]{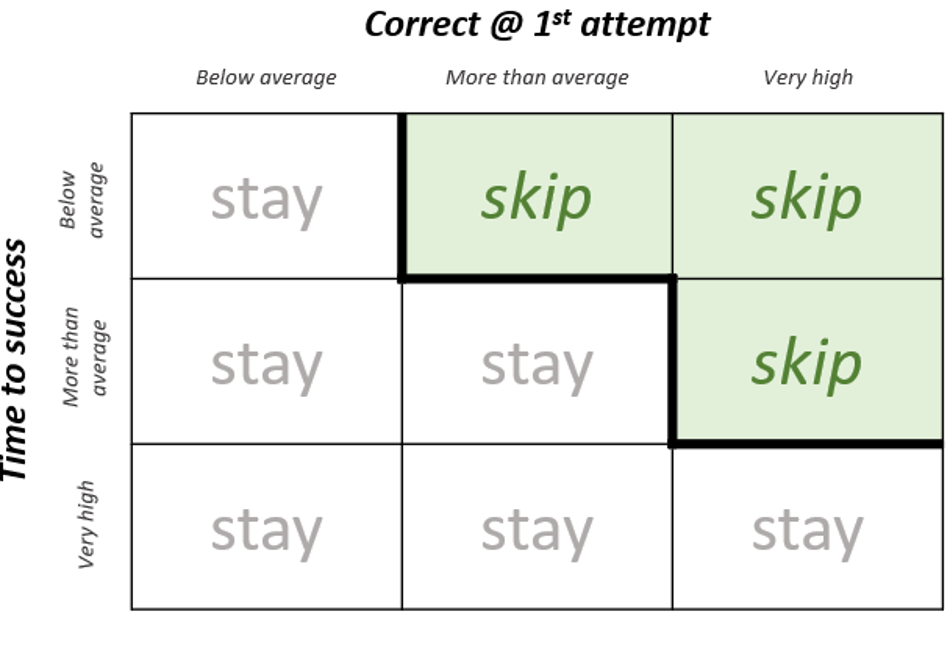}
    \caption{Non-bonus exercises.}
    \label{fig:ZPD-nonBonus}
  \end{subfigure}
  \hspace{2cm}
  \begin{subfigure}[b]{0.23\textwidth}
    \includegraphics[width=\textwidth]{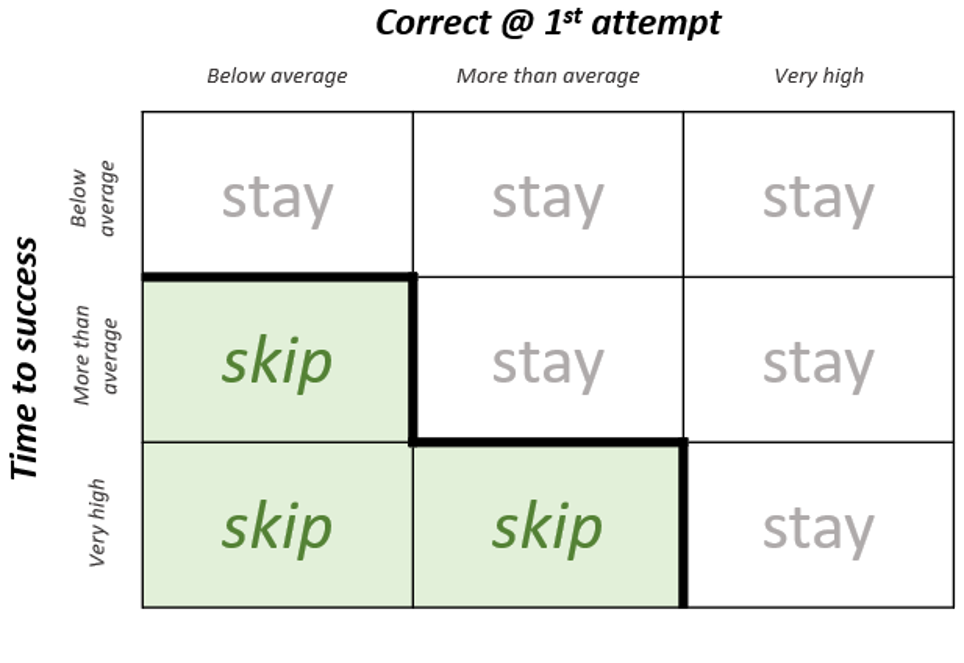}
    \caption{Bonus exercises}
    \label{fig:ZPD-Bonus}
  \end{subfigure}
 \caption{The interpretation of the Zone of Proximal Development by CFA and TTS predictions. (\protect\subref{fig:ZPD-nonBonus}) demonstrate the stipulated ZPD and the skipping policy for regular (non-bonus questions), while (\protect\subref{fig:ZPD-Bonus}) represents it for bonus questions.}
 \label{fig:ZPD-policy}
 \end{center}
\end{figure}

With these features in hand, machine learning models are trained to predict the CFA and TTS for each student for subsequent content. These predictions are then used to define a policy that is designed to keep students in their Zone of Proximal Development by ruling whether an exercise should be skipped. The skipping policy considers an exercise too easy for a student if she can solve it correctly and relatively fast. More precisely, if the CFA prediction model predicts that a student's probability of being correct is above a certain threshold and the TTS prediction model predicts that the time to solve this exercise will be below another threshold, then the policy rules the exercise as ``below" the ZPD, thus too easy, and recommends skipping it. Similarly, the policy will skip a bonus exercise that is too challenging by the predictive models, to prevent the overwhelming of struggling students (See Figure~\ref{fig:ZPD-policy}).

When deciding whether to skip an exercise or not, the predictive models use features that are derived from the experience of students that attempted to solve this exercise. This might become a problem once all students start using the adaptive engine since while using the engine, the students will only be present with an exercise that is not too easy nor too challenging for them. Therefore, this set of students is no longer a good representation of the population. To solve this problem, we borrow the idea of ``exploration" from Reinforcement Learning~\cite{sutton1998introduction}. Thus, even if the policy described above suggests that an exercise should be skipped, with a certain probability, the exercised will be served regardless as a way to collect data and improve future decisions.
The policy also respects signals from the content provider and teachers. For example, the content provider can flag an exercise as mandatory, in which case the exercise will be served to all students. This is useful, for example, when some exercises are used to grade students.

\section{Implementation}
The methods described above should be adapted to the specific nature of the e-learning system and the content it is delivering. In this section we describe the e-learning system we used and how we implemented E-gostky to work with this e-learning system.

\subsection{The Educational Content} \label{sec:content}
The content was provided by a major content provider\footnote{details are omitted for blind review} and was developed by pedagogical experts in the field. The goal of the program is to teach fractions to elementary school students in the $4^{\mbox{th}}$ grade. The program is built on the Action-Process-Object-Schema (APOS) theory. In this theory, students are provided with guidance and aids (video clips, hints, ...) that help students to develop the mental constructions needed to solve problems and gain a deeper understanding of the concept being learned.

The educational program includes 4 sections, each section is divided into topics and each topic covers several learning units. For each learning unit there are multiple pages, containing exercises and other educational material. Overall, the program is composed of 13 topics, 120 learning units, and 702 exercises. The content is delivered via a web-browser and students learn in class at their own pace. Each exercise is characterized by the learning topic it addresses (e.g., mixed numbers, adding fractions), its goal (such as concepts, implementation, creativity), its representation (pie, lines) and type (multi-choice, open question, cloze). The content provider did not make the information about the pedagogic dependencies between the topics and the learning units available for use with E-gotsky. Nevertheless, a learning path was provided, which is termed as the ``baseline path", designed by pedagogical experts to fit to the average learner.

In the beginning of every topic a didactic cartoon video was provided. Then, students practiced the topic by solving exercise and using embedded aids represents as text, video clips or interactive labs. Once an exercise is solved, after as many attempts as needed, the student is presented with the subsequent exercise or instructions. The students could also choose to check a ``I didn't understand" checkbox. In that case, the teacher was notified and helped these students individually or in small groups.
A quiz, that is administered at the end of each section, is a part of the content and is used to assess mastery levels. The quiz interface is similar to the exercises presented during the learning stage.

\subsection{Prediction Models and Skipping Policy} \label{sec:skippingPolicy}
180 different features were collected by the system. To reduce the complexity of the model, feature selection using Recursive Feature Elimination~\cite{guyon2002gene} with 5 fold cross-validation was used to select a subset of the features to be used at run time. We observed that reducing the number of features to 20 out of the 180 features available results in a minimal loss of accuracy. The features that were selected for each of the models, \textbf{CFA Prediction Model} and \textbf{TTS Prediction Model}, are listed in Table~\ref{tables:selectedFeatures}.
\begin{table}
\centering
\caption{Selected Features Groups}
\begin{tabular}{|p{0.2\columnwidth}|p{0.7\columnwidth}|} \hline
\bf{Model}&\bf{Features} \\ \hline
\bf{Correct at First Attempt (CFA) model}& the CFA of the last 1, 3, 5 and 20 previous exercises of the same type as the target exercise (4 features), the CFA of the last 20 previous exercises with the same representation as the target exercise (1 feature), TTS of the previous exercise (1 feature), target exercise mean CFA (1 feature), student's CFA and TTS in all previous exercises (2 features), mean CFA of target LU (over all exercises in that LU) (1 feature) \\ \hline
\bf{Time To Solve (TTS) model}& the student's TTS rank in the last 3, 10, 20 exercises compared to other students (3 features), student's TTS rank in the last 3, 5, 10, 20 exercises of the same type as the target exercise compared to other students (4 features), student's TTS rank in the 3 previous exercises with the same representation as the target exercise compared to other students (1 feature), student's TTS rank in the last 3 exercises with the same goals as the target exercise compared to other students (1 feature),exercise's STD duration (1 feature) \\ \hline
\end{tabular}
\label{tables:selectedFeatures}
\end{table}

With these features in hand, we used Random-Forests~\cite{breiman2001random} to train the models that predict the CFA and TTS for each student for subsequent exercise. 

The system requires some tuning such as selecting the features to use at run-time, training the models and selecting the thresholds for the skipping policy. This tuning was done using data that was collected during the previous school year when 714 students were learning using the same content but without the adaptive engine. We used this data also for offline validation of the skipping policy. The results of the evaluation are presented next.

\subsection{Offline Evaluation}
To evaluate our skipping policy, we compared it against two other policies: 
\begin{itemize}
    \item \textbf{Random} - Skip an exercise randomly with a fixed probability.
    \item \textbf{Benchmark} -	Skip after $n$ consecutive successes.
    \item \textbf{Adaptive} -	Use our skip policy as defined in Section~\ref{sec:methods}.
\end{itemize}
 Since the offline data contains students' performance on all the exercises we were able to compare the different policies using two main measures:
\begin{itemize}
    \item \textbf {False Positive Rate (FPR):} The percentage of exercises that the policy would have skipped but the student had hard time solving, that is, the student was not correct on the first attempt or took longer than average time to solve the exercise (``bad skips").
    \item \textbf {Time Saved:} How much time the policy saved, on average, for the students. This is the time it took the student to complete the exercises that were skipped divided by the time it took to complete all the exercises.
\end{itemize}
\begin{figure}
\centering
\includegraphics[width=\columnwidth]{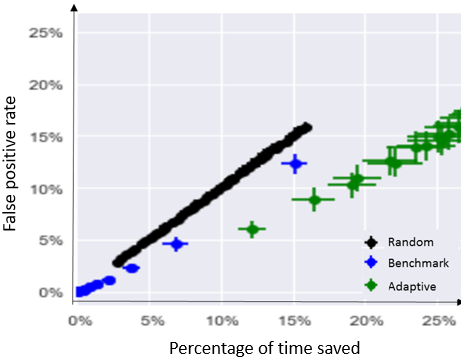}
\caption{Comparison of different skipping policies, with different parameters, on the offline data. The graph shows that the adaptive method dominates the other policies as it can save more time for a given false positive rate.\label{fig:policies}}
\end{figure}

The results of this comparison are presented in Figure~\ref{fig:policies}. The results show that the adaptive policy outperforms other policies since it can save more time for the student while making fewer false positive mistakes. 

These results were used to tune the parameters of the skipping policy. As Figure~\ref{fig:policies} demonstrates, these parameters allow controlling the balance between the risk of skipping over exercises that can be beneficial for the student, in terms of ZPD, and the benefit of saving time. In other word, these parameters allow tuning the boundaries of the predicted ZPD. Aggressively tuned parameters will result in more common or bigger skips, which in turn will cause presenting exercises that are too difficult. However, conservative parameters will err on the direction of presenting exercises that are too easy. In our implementation we selected the parameters such that the false positive rate will be below $10\%$ while saving as much time as possible.

\begin{figure}
\centering
\includegraphics[width=\columnwidth]{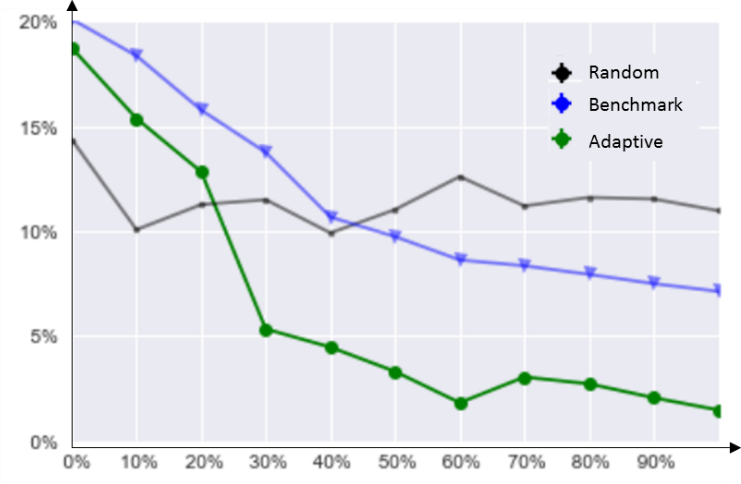}
\caption{The percentage of time saved (y-axis) vs the rank of the student in terms of rate of answering answers correctly at first attempt (x-axis). 
The plot presents values of different simulation results for each policy.
\label{fig:personalization}}
\end{figure}

Figure~\ref{fig:policies} shows that on average the adaptive policy can save time for the students while making few mistakes. However, this does not mean that personalization is happening. For example, it is possible that the content contains exercises that are easy for all students and the policy learned to skip these exercises for all the students. This behavior is not expected of the Random policy since the rate in which it skips exercises is independent of the student but it is not clear for the other policies.
To test this property we measured the time saved for each student and ranked the students based on their correctness rate in answering an exercise. The results of this evaluation are presented in Figure~\ref{fig:personalization}. As expected, the Random policy saves the same amount of time for all the students while the adaptive policy saves significant amount of time for the top $20\%$ of the students while providing the bottom $40\%$ of the students with more opportunities to practice their skills. Note that this works to the benefit of both the advanced and the struggling students. 

\subsection{Interface to the e-learning system}
The methods described above were implemented as a web-service. The content delivery system sends updates about students' progress and the adaptive learning engine, E-gostky, replies with the recommended next exercise to deliver. 
When a student solves an exercise, a request for the next exercise is sent from the content delivery system. E-gostky predicts the CFA and TTS for the 5 subsequent exercises as they appear in the baseline curriculum using the CFA and TTS prediction models, and decides whether to skip it or not according to certain thresholds as mentioned in Section~\ref{sec:skippingPolicy}. If none of the 5 exercises is recommended by the skipping policy, then the exercise that immediately follows them is selected. Otherwise, the first exercise that is recommended by the skipping policy is selected. 

\begin{figure}
\centering
\includegraphics[width=\columnwidth]{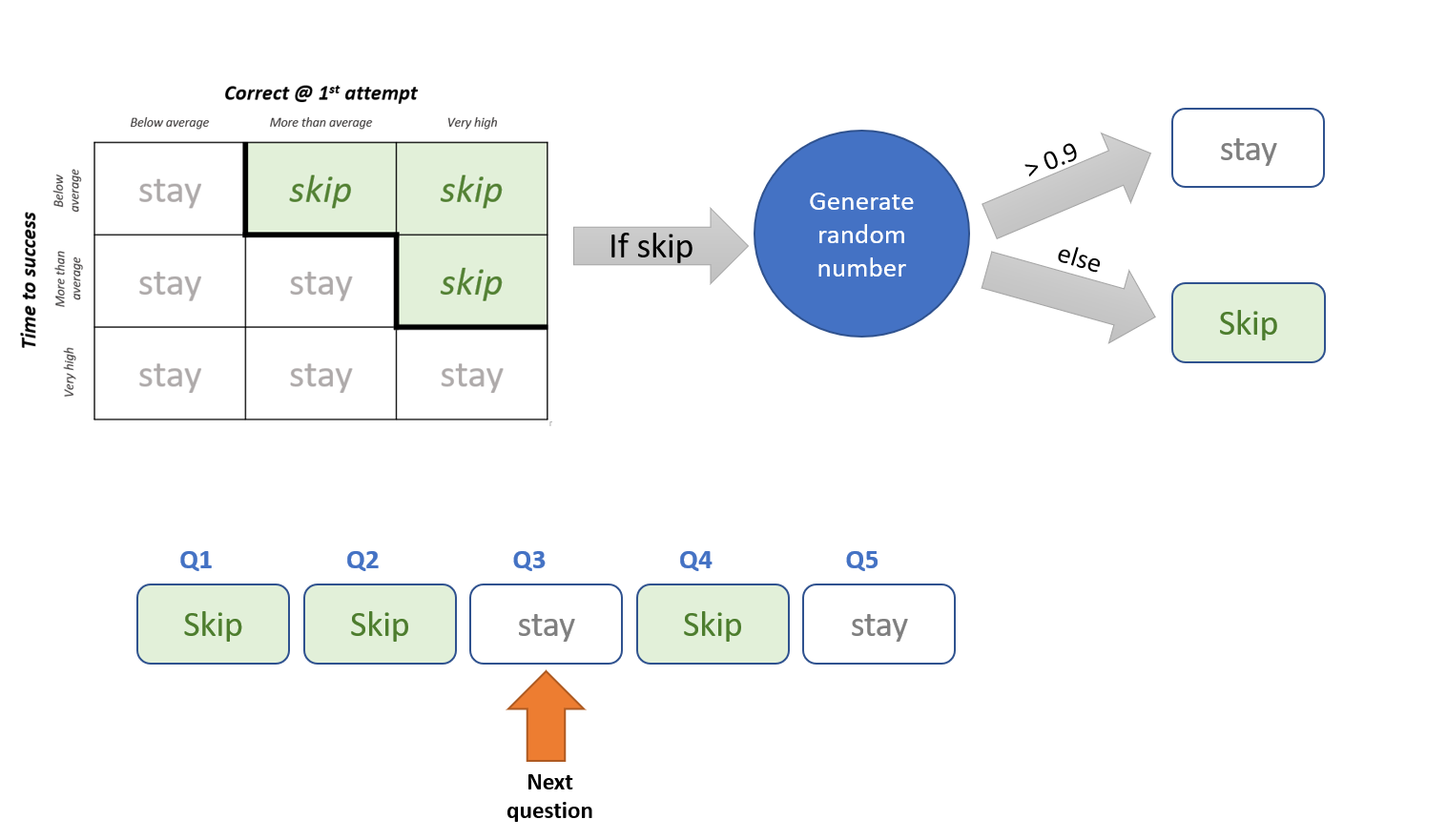}
\caption{Description of E-gostky pipeline. At first a decision is made for each exercise according to the predicted CFA and TTS. Then, given the decisions for the 5 candidates, the engine will recommend as the next exercise the first exercise that it has decided not to skip. \label{fig:pipeline}}
\end{figure}

Furthermore, we implemented the ``exploration" mechanism that forces an exercise to be delivered with a probability of $10\%$ even if the policy suggested that it should be skipped. We also allowed the content provider to mark exercises as mandatory in which case these exercises were delivered to all the students. In addition, we adjusted E-gostky to handle ``bad skips". If a student does not solve correctly at first attempt an exercise that followed a skip, E-gostky sends her backwards in the curriculum, to the exercise that would have been presented, if there were no skips. An illustration of the pipeline is presented in Figure~\ref{fig:pipeline}.

 The thresholds of the system and the number of candidate exercises to skip were empirically selected during the offline evaluation phase. The maximum number of successive exercises E-gostky can skip is set to 5. Also, mandatory questions cannot be skipped (even if the engine recommends skipping them).

The web-service also logs the information for analysis and for updating the features used by the predictive models. In our implementation these features were updated on a weekly cadence. It is important to note that no Personal Identifiable Information is available to the adaptive learning engine.

\section{In-class experiment}\label{sec:in class}
After verifying the system using offline data and tuning its parameters we evaluated the system in the classroom.
The experiment was conducted during the 2017-2018 school year in 18 classrooms from 8 schools. Overall, 412 students participated in the experiment. Unfortunately, in 4 of the 8 schools the data that was collected was not valid for the statistical analysis. In the first two schools the computer system was malfunctioning and in the others, the teachers did not follow the guidelines of the experiment and did not use the e-learning system to teach all the topics. This left 4 schools and 213 students for analysis purposes.

One of the pedagogical experts' requirements while designing the experiment was to allow all students to experience with the adaptive learning system for at least half of the sections. Therefore, we used mixed design for the experiment. Students were assigned, at random, to one of three groups. Each group was assigned to use either the adaptive engine or the baseline engine in each of the 4 sections of the curriculum (see Table~\ref{tables:experimentDesign}). 

\begin{table}
\centering
\caption{Classroom experiment design}
%\begin{tabular}{|P{1.5cm}|P{1.2cm}|P{1.2cm}|P{1.2cm}|P{1.2cm}|}\hline
\begin{tabular}{|c|c|c|c|c|c|c|} \hline
&\bf{Section 1}&\bf{Sec. 2}&\bf{Sec. 3}&\bf{Sec. 4} \\ \hline
\bf{Group A ($30\%$)}& ADL & ADL & ADL &ADL\\ \hline
\bf{Group B ($30\%$)}& ADL & ADL & BL &BL\\ \hline
\bf{Group C ($40\%$)}& BL & BL & ADL &ADL\\ \hline
\end{tabular}
\label{tables:experimentDesign}
\end{table}

The main analysis was done on the first two sections of the educational program where $60\%$ of the students were using E-gostky and the rest were using the curriculum as designed by pedagogical experts (baseline). We used the other sections to measure memory effects although the measurements on these sections are limited since not all classes and not all students in these classes completed these sections.
%Furthermore, since we didn't have benchmarks scores before starting the program, it is possible that the allocation of students to the different groups is biased. Therefore, our main measurements are comparing the progress students made during Section 2.

\section{Experiment Results}
In what follows, we compare the performance of students who studied using E-gostky, under the adaptive condition (ADL) against those who studied with the baseline path (BL) during the first two parts of the experiment. 
%We wished to control the class parameter to ensure that prior knowledge, level differences between the classes and teachers' effect will not influence the results. 

\begin{figure}
\centering
\includegraphics[width=\columnwidth]{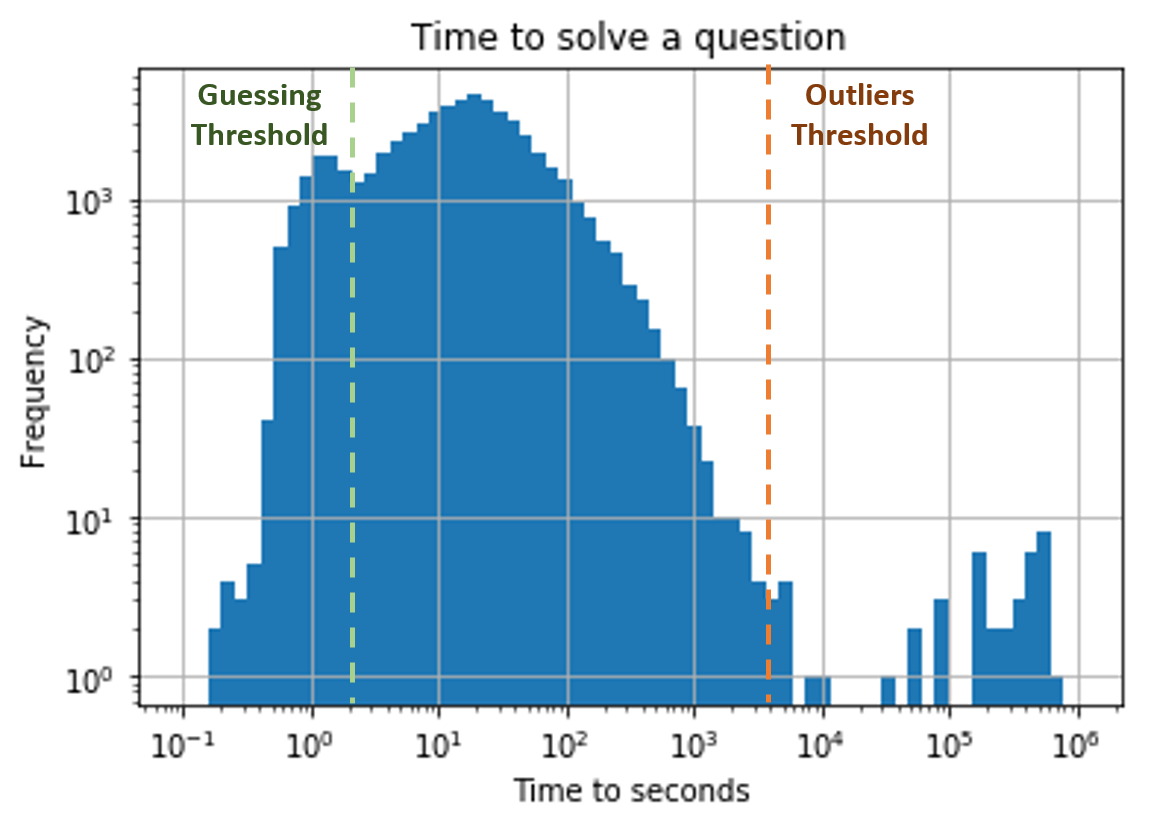}
\caption{TTS distribution and filtering thresholds. One threshold is for the guessing, where values below that threshold are considered a guess, and the other is for outliers, which all the observations above it are regarded as noise.
\label{fig:threshes}}
\end{figure}

\textbf{Total time spent.} 
The total time spent is the sum of the time that the student spent on all the exercises she was asked to solve. The data does not tell us if a student was actively trying to solve an exercise or was she distracted and therefore, we excluded from this computation all the exercises that took more than 25 minutes to complete (longer exercises are only $0.13\%$ of the observations, see Figure~\ref{fig:threshes}).

\begin{figure}
\centering
\includegraphics[width=0.95\columnwidth]{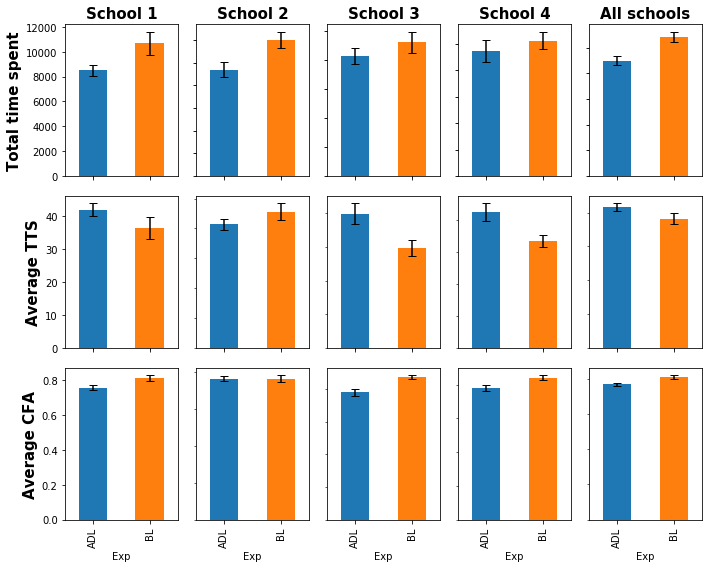}
\caption{A summary of the results. The first row represent the total time spent, the second represents the average time to solve an exercise and the third represents the mean success rate as measured by correct at first attempt, for non-quiz exercises. The results are presented for each school separately, and for all schools taken together. In each graph the left bar is for students who worked with the adaptive path and the right stands for the baseline. The graphs show that students in the adaptive path spent less time in total to complete the first two sections, despite the fact that they spent more time solving each exercise. Also, their success rate is lower which suggests they were more challenged than the students in the baseline path.
\label{fig:results1}}
\end{figure}

Students using E-gostky finished the exercises of the first two parts of the educational program $17\%$ faster than the students in the baseline path (see Figure~\ref{fig:results1} and Table~\ref{tables:TotalTime} for details). This difference was significant ($p$ = .025). To control the class' or teacher's effect, and since the assumptions for a two-way ANOVA test were not met, we compared both conditions in the following way: first, we compared the median time to answer each exercise per treatment per school using a Man-Whitney U two-tailed test. Then, we used Bonferroni correction to get the p-value mentioned above.

\begin{table}
\centering
\caption{Total time spent, in seconds, to complete the exercises of section 1 and 2 of the program. Median computed over all the students. Corrected p-val. for all schools = .025. }
\begin{tabular}{|c|c|c|c|c|c|c|} \hline
&\multicolumn{2}{c|}{\bf{Median}}&\multicolumn{2}{c|}{\bf{N}}&\multicolumn{2}{c|}{} \\
\hline
\bf{School}&\bf{ADL}&\bf{BL} &\bf{ADL} &\bf{BL} &\bf{U}  &\bf{p-val.} \\ \hline
\bf{1}& 9121 & 10973 & 33 & 22 & 264.0& 0.091\\ \hline
\bf{2}& 8536 & 11666 & 43 & 27 & 354.0 & 0.006\\ \hline
\bf{3}& 7285 & 9423 & 19 & 10 & 68.0 & 0.224 \\ \hline
\bf{4}& 8039 & 9749 & 30 & 19 & 224.0 & 0.214\\ \hline
\end{tabular}
\label{tables:TotalTime}
\end{table}

\textbf{Time to solve an exercise.}\label{sec:TTS_results} Comparing exercises that appeared in the learning stage (not in the quizzes), students using E-gostky spent more time on each exercise (see Figure~\ref{fig:results1} and Table~\ref{tables:TTS} for details). However, this effect was not significant ($p$=.086). In this analysis we dealt with outliers in the same way as in the previous analysis. Also, we discarded exercises that were answered in less than $2.1$ seconds, under the assumption that the answers to these exercises were guessed. We chose the threshold of guessing by visually inspecting the time to solve distribution as was suggested by~\cite{wise2006responsetime} and can be seen in Figure~\ref{fig:threshes}.
Since the ANOVA's assumptions were not met, we compared both conditions in the same way as in the analysis above. %Students in the adaptive path spent more time in average for each exercise, comparing to students in the baseline path.

\begin{table}
\centering
\caption{Median time to solve an exercise (in seconds). Corrected p-val. for all schools = .086.}
\begin{tabular}{|P{1cm}|P{0.8cm}|P{0.8cm}|P{0.65cm}|P{0.5cm}|P{0.8cm}|P{0.9cm}|} \hline
%\begin{tabular}{|c|c|c|c|c|c|c|} \hline
&\multicolumn{2}{c|}{\bf{Median}}&\multicolumn{2}{c|}{\bf{N}}&\multicolumn{2}{c|}{} \\
\hline
\bf{School}&\bf{ADL}&\bf{BL} &\bf{ADL} &\bf{BL} &\bf{U}  &\bf{p-val.} \\ \hline
\bf{1}& 42.049 & 38.503 & 35 & 23 & 301.0 & 0.108 \\ \hline
\bf{2}& 41.216 & 44.275 & 44 & 30 & 563.0 & 0.288\\ \hline
\bf{3}& 35.189 & 27.700 & 21 & 11 & 57.0 & 0.021\\ \hline
\bf{4}& 39.263 & 31.076 & 30 & 19 & 185.0 & 0.041 \\ \hline

\end{tabular}
\label{tables:TTS}
\end{table}

\textbf{Success Rate at First Attempt.} Since in this case too the assumptions for the ANOVA test were not met, we used Man-Whitney U with Bonferroni correction as before. The success rate of students in the adaptive path was significantly lower ($p$ = 0.010) than the baseline students' in non-quiz questions (see Figure~\ref{fig:results1}, and Table~\ref{tables:CFA} for details).

\begin{table}
\centering
\caption{Median success rate at solving an exercise correctly at the first attempt. Corrected p-val. for all schools=0.010.}
\begin{tabular}{|c|c|c|c|c|c|c|} \hline
&\multicolumn{2}{c|}{\bf{Median}}&\multicolumn{2}{c|}{\bf{N}}&\multicolumn{2}{c|}{} \\
\hline
\bf{School}&\bf{ADL}&\bf{BL} &\bf{ADL} &\bf{BL} &\bf{U}   &\bf{p-val.} \\ \hline
\bf{1}& 0.773 & 0.835 & 35 & 23 & 245.0 & 0.013 \\ \hline
\bf{2}& 0.776 & 0.779 & 44 & 30 & 658.0 & 0.987 \\ \hline
\bf{3}& 0.795 & 0.874 & 21 & 11 & 39.0 & 0.003\\ \hline
\bf{4}& 0.794 & 0.860 & 30 & 19 & 159.0 & 0.010 \\ \hline
\end{tabular}
\label{tables:CFA}
\end{table}
 
These last results were unsurprising since E-gostky was designed to skip exercises with a high probability to answer correctly and with low expected time to solve. This suggests that the engine worked as expected and challenged the students, moving them outside of their comfort zone, and into their Zone of Proximal Development.

\textbf{Correlation between number of exercises and quiz score.} The number of exercises the students were presented with was negatively correlated with success in quizzes for the adaptive group (Section 1: $r$ = -0.213, Section 2: $r$ = -0.07) and positively correlated with the quiz score of the baseline group (Section 1: $r$ = 0.218, Section 2: $r$= 0.330). With that said, the correlation between the number of exercises solved and quiz score was positive for both groups. This provides further evidence that the engine worked as planned and skipped more exercises for students that had a high mastery level.

\textbf{Quiz grades.} To investigate E-gostky's effect over the student's grades, we compared the students' average scores in the first two quizzes in two ways. %We only considered students who did the quiz, i.e., answered at least one exercise. 
First, since the data did not meet with the ANOVA's assumptions, we analyzed the differences using Man-Whitney U with Bonferroni correction again. For quiz 1, the grades of the students in the baseline group were higher, and this difference was borderline significant ($p$=.049). In the second quiz there were no significant differences ($p$=.386) (see Table~\ref{tables:Quizzes}). However, while inspecting the first quiz scores we found that there is one school in which the differences are indeed significant (School 4).

 Since the groups were assigned randomly, it might be possible that the BL and ADL groups were not balanced. Hence, we evaluated the improvement rate between the quizzes (in the same way as before as the ANOVA's assumptions were not met). The adaptive group's mean improvement were higher, but only borderline significantly ($p$=0.06). However, when considering only the school mentioned above (School 4), not only the adaptive group have significantly higher improvement rates ($p$ = 0.015), but also, their score in the second quiz was higher, compared to the baseline group as can be seen in Table~\ref{tables:Improvements}.

We also performed a paired t-test, where for each question we computed the probability to succeed in, for both group, for each school separately.
The probability to succeed was defined as the number of students who succeeded in an exercise out of the total number of students. In both quizzes, the baseline path students' grades were significantly higher ($p$ =.008 for the first quiz and $p$ =.002 for the second one).
However, when excluding the School 4 from the analysis, the grade differences for quiz 1 are not significant ($p$ =.108). This is not true for quiz 2 where in that case, the differences remain significant ($p$ = .008).

\begin{table*}
\centering
\caption{Quiz grades computed with a Man-Whitney U with a Bonferroni correction.
The grades are represented as the number of correctly answered questions out of the total number of questions (13 questions in quiz 1 and 19 in quiz 2).
The corrected p-value across all the schools in quiz 1 is 0.049 and for quiz 2 it is 0.386.}
\renewcommand{\arraystretch}{1.2}
\renewcommand{\tabcolsep}{4pt}
\begin{tabular}
{|l|l|c|c|c|c|c|c|c|c|c|c|c|c|c|c|c|c|}
\hline
&
&\multicolumn{4}{c|}{\bf{School 1}}
&\multicolumn{4}{c|}{\bf{School 2}}
&\multicolumn{4}{c|}{\bf{School 3}}
&\multicolumn{4}{c|}{\bf{School 4}} \\
\cline{3-18}
&&\bf{med.}&\bf{N}&\bf{U}&\bf{p.val}&\bf{med}&\bf{N}&\bf{U}&\bf{p.val}&\bf{med.}&\bf{N}&\bf{U}&\bf{p.val}&\bf{med.}&\bf{N}&\bf{U}&\bf{p.val}\\
\hline
\multirow{2}{*}{\bf{Q1}}&\bf{ADL}& \(\nicefrac {9}{13}\)&33&348&0.59&\(\nicefrac {7}{13}\)&41&536&0.64& 
\(\nicefrac {9}{13}\)&20&89&0.64&\(\nicefrac {9}{13}\)&30&165&0.01
\\ 
&\bf{BL}&\(\nicefrac {9}{13}\)&23&&&\(\nicefrac {8}{13}\)&28&&&\(\nicefrac {9}{13}\)&10&&&\(\nicefrac {10}{13}\)&19&&
\\[1pt]
\hline
\multirow{2}{*}{\bf{Q2}}&\bf{ADL}&\(\nicefrac {14}{19}\)&30&301&0.80&\(\nicefrac {10}{19}\)&33&256&0.42&\(\nicefrac {13}{19}\)&15&34&0.10&\(\nicefrac {15}{19}\)&30&254&0.79\\

&\bf{BL}&\(\nicefrac{14}{19}\)&21&&&\(\nicefrac {11.5}{19}\)&18&&&\(\nicefrac {15}{19}\)&8&&&\(\nicefrac {14}{19}\)&19&&\\[1pt]
\hline
\end{tabular}
\label{tables:Quizzes}
%\vspace{-1em}
\end{table*}

The fact that the improvement rates are slightly better for the treatment group (non significant) while the quiz scores are somewhat better for the control group may imply that the assignment of students to the treatment conditions was biased. Moreover, since some statistical test show the differences in the scores to be significant while other tests do not, suggests that the differences, even if existent, are not big and can controlled by a slight change the the thresholds of the skipping policy.
\begin{table}
\centering
\caption{Improvement from quiz 1 to quiz 2}
\begin{tabular}{|c|c|c|c|c|c|c|} \hline
&\multicolumn{2}{c|}{\bf{Median}}&\multicolumn{2}{c|}{\bf{N}}&\multicolumn{2}{c|}{} \\
\hline
\bf{School}&\bf{ADL}&\bf{BL} &\bf{ADL} &\bf{BL} &\bf{U}  &\bf{p-val.} \\ \hline
\bf{1}& 0.06 & 0.09 & 30 & 21 & 311.0 & 0.947 \\ \hline
\bf{2}& -0.01 & -0.06 & 31 & 17 & 229.5 & 0.470 \\ \hline
\bf{3}& -0.01 & 0.06 & 15 & 8 & 43.5 & 0.301 \\ \hline
\bf{4}& 0.12 & 0.05 & 28 & 19 & 153.0 & 0.015  \\\hline
\end{tabular}
\label{tables:Improvements}
%\vspace{-4em}
\end{table}

\begin{figure}[ht]
\centering
\includegraphics[width=0.9\columnwidth]{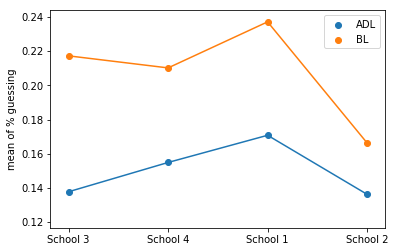}
\caption{Students' mean guessing rates.\label{fig:guessing}}
\vspace{1em}
\end{figure}

\textbf{Guessing rate.} Comparing the guessing rates as exercises that were answered in time that is lower than the threshold mentioned in Section~\ref{sec:TTS_results}, we found that students using E-gostky guess, on average, $25\%$ less compared to the students in the baseline path. In a 2-way ANOVA we found that both the treatment and the school had significant effects ($p$ < 0.0001 in both cases), while the interaction effect was non-significant as can be seen in Figure~\ref{fig:guessing}.

The lower guessing rates in the adaptive path indicates that the students using E-gostky were more engaged, as suggested by previous studies~\cite{joseph2005engagement}.

\subsection{Qualitative Results}
Feedback from the teachers and students was gathered by face-to-face meetings or by phone interviews.
The feedback reports reveal that the students enjoyed the overall experience of learning via a computerized environment. Besides that, they expressed special interest in working with the adaptive engine.

\textbf{Engagement and motivation.} The students reported that learning with E-gostky, which they nicknamed ``the skipping engine", was more engaging and motivating, as two of the students mentioned, ``I thought longer before answering each question, ... then I succeeded more and it skipped more questions for me...", ``it gave me motivation to continue and I started to be more accurate...". This behavior was noticed by teachers: ``There are children who came to class even in their free time", ``...
%this engine promoted the already existent `competition factor' that is in every activity of the students, but in a positive way. 
The student worked harder to get more skips than their friends...".

\textbf{Confidence and Self-efficacy} Another important aspect of E-gostky is the immediate feedback the student received. It made them more confident about their proficiency and future success. ``...It gave me a feeling that I know the material and that I can succeed each and every time", ``... it gave me more confidence as I was advancing and the feeling that I have a better understanding of the material".

\textbf{Usage of aids.} Another effect of E-gostky was that the students showed higher usage of the aids to increase their accuracy in solving exercises, to get more skips. As one of the students reported, ``...I used the lab more often and then I succeed more and then I got more skips".

\section{Discussion}
%The primary goal of the experiment was to validate that the adaptive learning engine, named E-gostky, does not harm student's learning while saving time.
The results show that students using E-gostky made similar progress between quizzes as did students in the control group. With that said, E-gostky indeed saved $\sim17\%$ of the time for the students that were using it (see Figure~\ref{fig:results1}). The time saved, which is about $\nicefrac{1}{6}$ of the time reserved for working with this e-learning system, can be used by teachers for other learning activities.

It might seem counter-intuitive at first glance that the students on the adaptive track spent more time on each exercise while their total time spent is smaller. This is explained by the fact that E-gostky skipped over many easy exercises that the student would have solved fast. As Vygotsky's theory suggests, these exercises do not contribute to the learning process since they are not in the ZPD and E-gostky learns to skip over these exercises. Another way to see that is the inverse correlation between the number of exercises a student answered and quiz results. For students with high mastery level many of the exercises are too easy and fall outside of their ZPD and therefore the adaptive engine does not present these exercises. In other words, while the common belief is that practice brings perfection, not every kind of practice improves mastery levels. 

We also see that students on the adaptive track were more challenged while solving exercises as can be seen on their lower success at first attempt rates (Figure~\ref{fig:results1}). This is another manifestation of the work done by E-gostky to eliminate too easy exercises. Vygotsky's theory suggests that selecting too hard exercises will result in frustration and disengagement but from the qualitative feedback we see that the system presented here did not suffer from this problem: teachers reported that students were eager to use the system even in their free time. Some teachers reported that even students with learning disabilities that typically cannot sit for the entire duration of the lesson found themselves engaged in learning. 

 One of the challenges of e-learning systems is the ``gaming the system" strategy students use, which is associated with reduced learning and performance~\cite{arbreton1998student, baker2004off,beck1997using,schofield1995computers}. As described by Baker et al. ``gaming the system" refers to behavior aimed at obtaining correct answers and advancing within the tutoring curriculum by taking advantage of regularities in the software instead of actively thinking about the material~\cite{baker2004off}. When using our approach, the positive feedback dissuades students from this behavior (prevents them from receiving skips), and motivates them to be accurate and as a result further enhances learning~\cite{baker2004off,beck1997using}.

There are some limitations to the experiment we conducted. We see differences in the scores achieved in the first quiz between the groups. These differences are influenced by one particular school in which the differences are significant. However, we learned, by interviewing the teachers after the experiment, that some of them did not fully understand the adaptive mechanism and thought that exercises were skipped by mistake. This confusion was also reported in students' feedback: ``At first I thought it was a bug in the software, so I went backwards to do skipped questions...", ``In the beginning I thought there was something wrong with the system, so I went back to the questions it skipped...". Also, supporting this explanation of the scores' differences, is the fact that in the second quiz, we found no significant differences, in this school and overall. With that said, in another set of tests we found significant differences between the groups. Since some tests suggest that the differences are significant and others not, we can conclude that even if differences exist, they are not big. It is possible that the tuning of the engine should be more conservative, with lower rate of skipping, to avoid this consequence. 

The original study design included 4 parts such that in the later parts students switch between the adaptive and the baseline path we did not use the later parts in the analysis. A significant fraction of the classes did not finish the later parts and from the little data we collected it looks like there are strong memory effects.

It is hard to distinguish between the contribution of E-gostky and the contribution of the notification students received about skipped exercises. Students reported that ``I thought longer before answering each question, ... then I succeeded more and it skipped more questions for me...", ``it gave me motivation to continue and I started to be more accurate...". Therefore, students learned that by providing accurate answers the system will skip over exercises and this was considered as a positive feedback. 
We conjecture that both factors contributed to the positive results of our system. However, quantifying the contribution is beyond the scope of this study.

\section{Conclusions}
We developed an adaptive learning engine, E-gostky, that personalize the learning curriculum by dynamically assessing the learning potential of the student at the current time and present content accordingly. Our approach relies on Vygotsky's Zone of Proximal Development theory. We quantified this notion into models that can be evaluated in practice, in real-time. 

E-gostky was incorporated in an e-learning system for fractions and deployed in 8 schools. We compared it to a baseline path which was built by domain experts. Using E-gostky saved $17\%$, which is \(\nicefrac{1}{6}\) of the students' time while maintaining similar learning outcomes. Moreover, the quantitative and qualitative results indicate that students were more engaged when using E-goskty.

The feedback provided by teachers and students show that the success of the system is not only due to its ability to select the right content but also due to the immediate feedback it provided. While traditional e-learning system only indicate whether a solution is correct or not, our system also presented the number of exercises skipped. This was rightfully interpreted by many students as a positive feedback and they quickly learned that in order to receive this feedback they have to try hard to solve exercises correctly and avoid guessing. As a result, students spent more time thinking about their solutions, and use more aids provided by the e-learning system.
Therefore, we conclude that when considering adaptive learning engines it is importance to pay attention not only to the content sequencing but also to the feedback students receive.

The setting in which we conducted the experiment allowed E-goskty to only skip over some exercises. Despite these limitations the results are very encouraging. However, it may be possible to extend it to other forms of differentiated instruction~\cite{hall2002differentiated} and provide even greater gains to students and teachers. These challenges are left for further study.
\bibliographystyle{abbrv}
\bibliography{adl} %
\end{document}